\documentstyle[12pt]{article}
\begin{document}

\title{Simply Modeling Meson HQET}

\author{B. Holdom\thanks{holdom@gpu.utcs.utoronto.ca} \hspace{0.5em}
and \hspace{0.5em}
M. Sutherland\thanks{marks@medb.physics.utoronto.ca}
     \and
      {\it Department of Physics, University of Toronto} \\
        {\it 60 St. George St., Toronto, Ontario, Canada M5S 1A7}  }
\date{UTPT-92-16; hep-ph/9211226}
\maketitle

\begin{abstract}
A simple relativistic model of heavy-quark-light-quark mesons is
proposed.  In an expansion in inverse powers of the heavy
quark mass we find that all zeroth and first order heavy quark
symmetry relations are satisfied.  The main results are:

$\bullet$ the difference between the meson mass and the heavy quark
mass plays a significant role even at zeroth order;

$\bullet$ the slope of the Isgur-Wise function at the zero recoil point is
typically less than $-1$;

$\bullet$ the first order correction to the pseudoscalar decay constant is
large and negative;

$\bullet$ the four universal functions describing the first order
corrections to the semileptonic decay form factors are small;

$\bullet$ these latter corrections are quite insensitive to the choice of
model parameters, and in particular to the effects of hyperfine mass
splitting.
\end{abstract}
\vspace{2ex}

\noindent {\bf 0.  Introduction}
\vspace{2ex}

It has recently been realized that major simplifications occur in
the theoretical treatment of weak decays of mesons containing a heavy
quark.\cite{1}  New symmetries appear in an expansion in inverse powers
of the heavy quark mass.  For example at lowest order the six form factors
for the decays of pseudoscalar to pseudoscalar and vector mesons become
related to a single universal form factor (the Isgur-Wise function).  The
latter depends only on the dynamics of the light degrees of freedom.  In
addition the masses of pseudoscalar and vector mesons with the same
heavy/light flavor content are equal at this order, as are their decay
constants.

However in the real world it is important to know the magnitude of the
$1/{m}_{Q}$ corrections.  It is expected that the $b$ quark is probably
heavy enough for the corrections to be small, but it is less clear for the
charm quark.  It is also becoming apparent that the size of the corrections
depends strongly on what quantity is being calculated.  We will attempt to
shed more light on these questions.

We will present a model based on simple finite quark loop graphs.  It is
relativistic
and incorporates the effects of hadronic recoil in a natural way.  We
demonstrate
explicitly that all heavy quark symmetry relations among the semileptonic decay
form
factors at zeroth and first order in $1/{m}_{Q}$ are satisfied by the model.
The
model also allows calculations to all orders in $1/{m}_{Q}$.

We consider our approach to be complementary to QCD sum rules.\cite{3}
The latter approach relies on a certain representation of the hadronic
contribution to a 3-point function.  The basic input for our model is a
representation of a Bethe-Salpeter amplitude for the heavy-quark-light-
quark meson.  As such, it may be improved by future
QCD-based numerical computations of Bethe-Salpeter amplitudes.  Our
model is also complementary to nonrelativistic quark models,\cite{14,2}
and it provides the first indication of how the $1/{m}_{Q}$ corrections are
affected by hyperfine mass splitting.  (Note that hyperfine mass splitting
effects are not included in the ISGW model.\cite{14})

Our model highlights the importance of the difference between a meson
mass and the mass of the associated heavy quark.\cite{4}  This mass
difference is needed to obtain consistent results at zeroth order in
$1/{m}_{Q}$ ({\it e.g.} the Isgur-Wise function).  We will show that
ignoring this difference leads to very different results.  And
correspondingly the $1/{m}_{Q}$ corrections to this mass difference must
be properly incorporated into the $1/{m}_{Q}$ corrections of other
quantities.

Another result is a confirmation that the $1/{m}_{Q}$ correction
to the $B$ meson decay constant ${f}_{B}$ is very large and negative.  This
was first suggested by lattice calculations,\cite{5} and it has been noted
in two dimensional calculations\cite{6} and sum rule
calculations\cite{7,8}.  This is to be contrasted with the corrections to the
weak
decay form factors, which we find to be relatively small.  The latter
observation also agrees qualitatively with sum rule results, although the
actual numerical values of various $1/{m}_{Q}$ corrections differ.

We test the sensitivity of our results to the parameters of the model, and
in particular we find that the form factor corrections are quite
insensitive.  We feel that the quantitative differences between our model
and sum rules give an indication as to the true uncertainty in present
theoretical determinations of these quantities.
\vspace{2ex}

\noindent {\bf 1. Definition of the model}
\vspace{2ex}

Before beginning we make the standard definitions of physical quantities.  The
meson
decay constants are given by
\begin{eqnarray} \left\langle{0\left|{{A}_{ \mu
}}\right|P(p)}\right\rangle =  i{f}_{P}{p}_{ \mu } & {\rm and} &
\left\langle{0\left|{{V}_{ \mu }}\right|V(p)}\right\rangle =
{f}_{V}{M}_{V}{ \varepsilon }_{ \mu }. \label{eq1and2} \end{eqnarray}
Our normalization is such that ${A}_{ \mu }=\overline{q}{ \gamma }_{ \mu
}{{ \gamma }_{5}Q}$.  The various form factors for semileptonic decay are
defined by
\begin{equation}
\left\langle{{P}_{2}({ v}_{2})\left|{{V}_{ \mu }}\right|{P}_{1}({
v}_{1})}\right\rangle =  \sqrt {{M}_{{P}_{1}}{M}_{{P}_{2}}}\left[{{h}_{
+}( \omega   ){({ v}_{1} +  { v}_{2})}_{ \mu } +  {h}_{ -}( \omega   ){({
v}_{1}
-  { v}_{2})}_{ \mu }}\right],
\label{eq3}
\end{equation}
\begin{equation}
\left\langle{{V}_{2}({ v}_{2})\left|{{V}_{ \mu }}\right|{P}_{1}({
v}_{1})}\right\rangle =  \sqrt {{M}_{{P}_{1}}{M}_{{V}_{2}}}{h}_{V}(
\omega   ){ \varepsilon }_{ \mu \upsilon \rho \sigma } {\varepsilon }_{
2}^{\ast \upsilon }{ v}_{  2}^{\rho }  { v}_{1}^{ \sigma },
\label{eq4}
\end{equation}
\begin{eqnarray}
\lefteqn{ \left\langle{{V}_{2}({ v}_{2})\left|{{A}_{ \mu
}}\right|{P}_{1}({ v}_{1})}\right\rangle = } \nonumber \\
 && -i\sqrt {{M}_{{P}_{1}}{M}_{{V}_{2}}}\left[{( \omega +  1){h}_{{A}_{1}}(
\omega   ){
\varepsilon }_{2 \mu }^{ \ast } -  \left({{h}_{{A}_{2}}( \omega   ){ v}_{1
\mu } +  {h}_{{A}_{3}}( \omega   ){ v}_{2 \mu }}\right) {\varepsilon }_{
2}^{\ast }\cdot { v}_{  1}}\right],
\label{eq5}
\end{eqnarray}
where the $ v$'s are meson velocities, and $\omega={v}_{1}\cdot{v}_{
2}$.

We choose to represent transition amplitudes by diagrams with
heavy mesons attached to a loop involving heavy and light quarks.  The
essential nonperturbative physics of QCD is that
contributions with large momentum flowing through the light quark line
are suppressed.  We will model this physics by including factors
in  the vertices which damp the loop integrals when the light quark
momentum $k$ is larger than some scale $\Lambda$ set by QCD.  This has two
desirable effects:  the integrals are made finite and we may consider
$k$ small in comparison to a meson momentum.
This latter fact is the crucial ingredient that gives rise to the correct
heavy quark symmetry relations.

The effective vertices between a light quark, a heavy quark and a
pseudoscalar or vector meson are taken to be
\begin{eqnarray} {\left({{\frac{{ Z }_{P}^{2}}{ -{  k}^{  2}+{\Lambda
}_{ -}^{  2}}}}\right)}^{n}{ \gamma }_{ 5} & {\rm and} & -i{\left({{\frac{{
Z }_{V}^{2}}{ -{  k}^{  2}+{\Lambda }_{+}^{  2}}}}\right)}^{n}{ \gamma
}_{ \mu }, \label{vertices} \end{eqnarray}
where $k$ is the momentum of the light quark.
The light and heavy quarks are assigned standard propagators
with masses ${m}_{q}$ and ${m}_{Q}$ respectively.
With these choices, and with $n$ an integer or half-integer, standard
methods involving Feynman
parameters may be used in the one loop computations.

We may now consider expanding various quantities to first order in $
\Lambda   /{m}_{Q}$.
\begin{equation} {\Lambda }_{\pm }=\Lambda \left({1+(\pm h-
g){\frac{\Lambda }{{m}_{Q}}}}\right) \label{eq6} \end{equation}
\begin{equation} { Z }_{P,V}   =A \Lambda {\left({{\frac{{  m}_{
Q}}{\Lambda }}}\right)}^{{\frac{  1}{4n}}}  \left({1+{B}_{P,V}{\frac{
\Lambda }{{  m}_{  Q}}}}\right) \label{eq7} \end{equation}
\begin{equation} {M}_{P,V}^{2}-{m}_{Q}^{2} =  c \Lambda {  m}_{  Q}
\left({1+{d}_{P,V}{\frac{ \Lambda }{{  m}_{  Q}}}}\right) \label{eq8}
\end{equation}
\begin{equation} {f}_{P,V}=a{\frac{{ \Lambda
}^{3/2}}{{{m}_{Q}}^{1/2}}}\left({1+{b}_{P,V}{\frac{ \Lambda }{{  m}_{
Q}}}}\right) \label{eq9} \end{equation}
The $P \leftrightarrow   V$ symmetry in (\ref{eq7}-\ref{eq9}) at zeroth
order is a direct consequence of the model, as shown in Appendix A. The
same is true
of the nontrivial zeroth order ${m}_{Q}$ dependence, and in particular the
standard scaling ${f}_{P,V} \propto   {{m}_{Q}}^{-1/2}$.  The zeroth
order constants $A$, $c$ and $a$ will be completely determined in terms
of $\Lambda/{m}_{q}$ and $n$ without further input.

The first order constants ${B}_{P,V}$, ${d}_{P,V}$ and ${b}_{P,V}$ depend
in addition on the values of $g$ and $h$.  These latter constants model the
effect of hyperfine mass splitting.  Through their effect on ${d}_{P,V}$
they determine how the pseudoscalar and vector masses approach a
common value in the heavy quark limit.  To first order in
$1/{m}_{Q}$ we will find that $g$ and $h$ dependence cancels out of
certain physical quantities.  As with $A$, $c$ and $a$, we stress that the
quantities ${B}_{P,V}$, ${d}_{P,V}$ and ${b}_{P,V}$ are not parameters of
the model and will be determined by $\Lambda/{m}_{q}$, $n$, $g$ and $h$.

For certain physical quantities we will need to know ${m}_{q}$ and $\Lambda$
separately.  It is reasonable that the appropriate
effective light quark mass should be of order its constituent mass, in
common with most successful quark models.  As for
$\Lambda$ we note the following relation which follows from (\ref{eq8})
\begin{equation} {\left.{{M}_{P,V}}\right|}_{\rm zeroth \; order}-{m}_{Q}
\equiv
\overline{ \Lambda }  ={\frac{ c}{2}}\Lambda. \label{eq12}
\end{equation}
The first equality is the standard definition of $\overline{ \Lambda }$
appearing in
the literature.  When necessary to choose $\Lambda$, it will chosen so as to
give reasonable values of $\overline{ \Lambda }$.  The parameter $n$
determines the form and extent of the damping due to
the meson vertex factors, and one of our goals will be to study the
sensitivity of $1/{m}_{Q}$ corrections to the choice of vertex factors.

The various weak decay form factors may also be expanded,
\begin{equation}
h_{i}(\omega ) = \alpha_{i}\xi(\omega ) +
            \delta_{c}h_{i}(\omega )\frac{\Lambda}{m_{c}}
	+ \delta_{b}h_{i}(\omega )\frac{\Lambda}{m_{b}}  \label{eq10}
\end{equation}
with ${ \alpha }_{+} ={\alpha }_{  V}={\alpha }_{{  A}_{  1}}={\alpha }_{{
A}_{  3}}= 1$ and  ${ \alpha }_{-} ={\alpha }_{{  A}_{  2}}= 0$.  $ \xi   (
\omega   )$ is the Isgur-Wise function and we will calculate it in terms of
$ \Lambda   /{m}_{q}$ and $n$.  In QCD it has been shown that numerous
relations exist between the first order corrections.\cite{4,9}  We demonstrate
in
Appendix B that these relations are satisfied by our model for any values of
the
parameters.  This consistency with heavy quark symmetry at first order in
$\Lambda/{m}_{Q}$ is a nontrivial test for any quark model of heavy mesons.

We will present numerical results for the first order corrections and their
first
derivatives, all at $\omega =1$.  We do not treat the perturbative QCD
corrections;
they have been calculated elsewhere\cite{13} and should be added to our
results.
\vspace{2ex}

\noindent {\bf 2.  Zeroth order results}
\vspace{2ex}

We first consider the results of our model at zeroth order in
$\Lambda/{m}_{Q}$.  Details of these calculations may be found in Appendix
A.  Computation of $c$ involves finding the zero of the meson two-point
function as given in (\ref{eqA4}).  Some examples of these ``mass
functions" are displayed in Fig. 1.
When we set the location of the zeros
equal to ${M}^{2} =  {m}_{Q}^{2} +  c \Lambda {  m}_{  Q}$ we find the
following values of $c$:
\begin{equation} \begin{array}{c|ccc} \Lambda /{  m}_{  q}&
{c}_{n=1}&{c}_{n=3/2}&{c}_{n=2}\\ \hline
2&1.5797&1.3949&1.3006\\
4&1.4635&1.1976&1.0545\end{array} \label{table1} \end{equation}
 We see that increasing $n$ decreases $M$.  $
\Lambda $ is constrained to be greater than ${m}_{q}$, since otherwise we
find no sensible zero of the mass function.  We will see later that the two
choices shown in (\ref{table1}), $ \Lambda   /{m}_{q}=2$ and 4,
imply a reasonable range for $\overline{ \Lambda }$.

 We display in (\ref{table2}) the results for the zeroth
order
parameters $A$ and $a$ appearing in ${ Z }_{P,V}$ and ${f}_{P,V}$
respectively and given by (\ref{eqA7}) and (\ref{eqA10}).
\begin{equation} \begin{array}{c|ccc|cc} \Lambda /{  m}_{  q}&
{A}_{n=1}&{A}_{n=3/2}&{A}_{n=2}&a_{n=3/2}&a_{n=2} \\  \hline
2&0.9181&0.9184&0.9059&0.25&0.17 \\
4&1.1461&1.1358&1.1077&0.29&0.19 \end{array} \label{table2} \end{equation}
We do not include in (\ref{table2}) results for $a_{n=1}$ because in this case
the dependence of  ${f}_{P}$  on $ \Lambda $ is not of the form of
(\ref{eq9}) (it is logarithmic), while ${f}_{V}$ diverges.  For $n=3/2$
${m}_{b}=4.8$ GeV and $\Lambda=500$ MeV and 1 GeV we find
${f}_{B}={f}_{{B}^{{}^*}}=40$ MeV and 132 MeV respectively at zeroth
order.  The strong $\Lambda $ dependence reflects ${f}_{P,V}\propto
{\Lambda }^{3/2}$ at zeroth order.  ${f}_{P,V}$ become smaller for larger
$n$.

The Isgur-Wise function $ \xi (\omega )$ of our model is given by
(\ref{eqA18}).  It depends only on the index $n$ and the ratio $ \alpha
\equiv   {m}_{q} /\Lambda $.  Let us see the effect of ignoring the zeroth
order mass difference $\overline{ \Lambda }$.  Putting $c=0$ in
(\ref{eqA18}) allows one to perform the relevant integrations
explicitly, yielding a linear combination
\begin{equation} { \xi }^{c=0}( \omega   )={\frac{{f}_{1}( \alpha   ){ \xi
}_{1}( \omega   )+{f}_{2}( \alpha   ){ \xi }_{2}( \omega   )}{{f}_{1}( \alpha
){ \xi }_{1}( 1 )+{f}_{2}( \alpha   ){ \xi }_{2}( 1 )}} \label{eq15}
\end{equation}
of two functions independent of $n$ and $ \alpha $,
\begin{eqnarray} { \xi }_{1}( \omega   )={\frac{2}{1+ \omega }} &
{\rm and} & { \xi }_{2}( \omega   )={\frac{ \ln\left({\omega   +\sqrt {{
\omega }^{ 2}- 1}}\right)}{\sqrt {{\omega }^{2} -1}}} .\label{eq16}
\end{eqnarray}
The coefficients are given by
\begin{eqnarray} {f}_{1}( \alpha   )={\left.{{\left({{\frac{ \partial
}{\partial   t}}}\right)}^{2n- 1}{\frac{ \pi }{\sqrt {  t}  + \alpha
}}}\right|}_{t=1} & {\rm and} & {f}_{2}( \alpha   )={\left.{{\left({{\frac{
\partial }{\partial   t}}}\right)}^{2n- 1}{\frac{ \alpha    \ln (t /{\alpha
}^{2}
)}{t-{ \alpha }^{ 2}}}}\right|}_{t=1} \label{eq17} \end{eqnarray}

The expression (\ref{eqA18}) for $ \xi (\omega )$ with nonzero $c$
must be evaluated numerically.  We plot the result in Fig. 2  together with
the $c=0$ form in (\ref{eq15}).  The difference is quite significant; this
demonstrates the importance of retaining the zeroth order mass
difference $\overline{ \Lambda }$.  A simple quark loop model was
recently described by deRafael and Taron\cite{10} in which they obtained
$ \xi
(\omega )={\xi }_{  2}  ( \omega   )$.  But that model, like our model with
$c=0$, does not correctly take into account a nonzero $\overline{ \Lambda
}$.

It is interesting to see how the $ \xi (\omega )$ with nonzero $c$ depends
on $ \alpha $.  We plot in Fig. 3 the first and second derivatives of $ \xi
(\omega )$ at $ \omega   =1$, for the case $n=3 /  2$.  We see that there is
a range of $ \alpha $ over which they are quite insensitive to the value of
$ \alpha $ .  Our two choices $ \alpha   =1/2$ and $ \alpha   =1/4$ more or
less fall within the region of insensitivity.

Although we cannot obtain $ \xi (\omega )$ analytically we have checked
numerically that it has the correct behavior at large $ \omega $, namely it
tends to zero.  We have also computed the derivatives of $ \xi (\omega )$
at $ \omega   =1$ :
\begin{equation} \begin{array}{c|ccccc}{\frac{ \Lambda }{{  m}_{  q}}}
,n& \xi   (1)&{ \xi }^{ '}(1)&{\frac{{ \xi }^{ ''}(1)}{2}}&{\frac{{
\xi }^{ '''}(1)}{3!}}&{\frac{{ \xi }^{ ''''}(1)}{4!}}\\  \hline
2,{\frac{3}{2}}&1&
-1.241&1.370&-1.459&1.536\\4,
{\frac{3}{2}}&1&-1.118&1.047&-0.917&0.777\\2,1&1&-
1.359&1.691&-2.094&2.631\end{array} \label{table4} \end{equation}
Our values for ${ \xi }^{ '}(1)$ are
consistent with the present experimental de\-ter\-mi\-na\-tions.\cite{12}
(Note that
these determinations assume some functional form
for $\xi(\omega)$ which is not identical to ours.)  We see that
our model, like most other models, does not satisfy the constraints on ${
\xi }^{ '}(1)$ and ${ \xi }^{ ''}(1)$ recently argued for in \cite{10}.  It is
amusing that the constraints {\it are} satisfied by the inconsistent
$c=0$ model.

\vspace{2ex}
\noindent {\bf 3.  Order-$ \Lambda /{  m}_{  Q}$ corrections}

\vspace{1ex}

The purpose of this section is to find the $O(\Lambda /{m}_{Q})$
corrections and to explore their dependence on the parameters, $n$,
$\Lambda $, $g$, and $h$.  For $(\Lambda /{m}_{q},n)$ we take the three
sets of values $(2,{\frac{3}{2}})$, $(4,{\frac{3}{2}})$, and $(2,1)$.  And for
each quantity we calculate we will indicate explicitly the $g$ and $h$
dependence.

We remark that just as the zeroth order $ \Lambda $ dependence of the
meson masses ({\it i.e.}  $c$) plays an important role in the zeroth order
results, the next to leading $ \Lambda $ dependence of the meson masses
($i.e.$ ${d}_{P,V}$) plays an important role in the first order corrections.
The meson masses enter the loop calculations via the on-shell external
momenta ${p}^{2} =  {M}_{P,V}^{2}$.  $ \Lambda $ dependence from meson
masses also originates in the factor ${M}_{V}$ in the definition of the
vector decay constant (\ref{eq1and2}) and in the factors $\sqrt
{{M}_{{P}_{1}}{M}_{{P}_{2}}}$ and  $\sqrt
{{M}_{{P}_{1}}{M}_{{V}_{2}}}$ in
the definition of the form factors (\ref{eq3}-\ref{eq5}).

Our results for the corrected masses ${M}_{P,V}$ and the corrected vertex
normalizations ${ Z }_{P,V}$  ({\it i.e.} the parameters ${d}_{P,V}$
and ${B}_{P,V}$) are
\begin{equation} \begin{array}{c|cc}
\Lambda /{  m}_{  q},n&{  d}_{  P}&{  d}_{  V}\\  \hline
2,{\frac{3}{2}}&0.379-(h+g)0.646&0.335+(h-g)0.646\\
4,{\frac{3}{2}}&0.385-(h+g)0.886&0.262+(h-g)0.886\\
2,1&0.461-(h+g)0.801&0.365+(h-g)0.801\end{array} \label{table6}
\end{equation}

\begin{equation} \begin{array}{c|cc}
\Lambda /{  m}_{  q},n& {  B}_{  P}&{  B}_{  V}\\  \hline
2,{\frac{3}{2}}&0.266-(h+g)1.387&0.301+(h-g)1.387\\
4,{\frac{3}{2}}&0.193-(h+g)0.989&0.274+(h-g)0.989\\
2,1&0.404-(h+g)1.341&0.535+(h-g)1.341\end{array}
\label{table7} \end{equation}

The values of the correction coefficients for the decay constants
as defined in (\ref{eq9}) are displayed in (\ref{table8}).
Note in particular the large negative values of
$b_{P}$ for the pseudoscalar.  (From our parameter fit at the end we find
that both $g$ and $h$ are positive.)
The implication is that the first order corrections to
${f}_{B}$ are nearly of the same order (with opposite sign) as the zeroth
order values.  We conclude that the $ \Lambda /{  m}_{  Q}$ expansion for
${f}_{B}$  is breaking down.
\begin{equation} \begin{array}{c|cc}
\Lambda /{  m}_{  q},n&{  b}_{  P}&{  b}_{  V}\\  \hline
2,{\frac{3}{2}}&-2.92-(h+g)1.82&-1.37+(h-g)1.82\\
4,{\frac{3}{2}}&-3.82-(h+g)1.52&-1.43+(h-g)1.52\end{array}
\label{table8} \end{equation}

We now turn to the first order corrections to the weak decay form factors.  It
is
standard\cite{3} to write the first order terms in (\ref{eq10}) as
\begin{equation}
{\gamma }_{i}(\omega )\xi (\omega ) \equiv
{\delta }_{c}{h}_{i}(\omega ){\frac{\Lambda }{{m}_{c}}}+{\delta
}_{b}{h}_{i}(\omega ){\frac{\Lambda }{{m}_{b}}}.
\label{c1} \end{equation}
In QCD the first order terms may be expressed in terms of the universal
functions ${
\chi }_{1,2,3}( \omega   )$ and ${ \xi }_{3}( \omega   )$ in addition to $ \xi
(\omega
)$.\cite{4,9}  These relations are reproduced in Appendix B.  Also in Appendix
B we
demonstrate that these relations are satisfied by our model for any values of
the
parameters.  We also verify in detail Luke's Theorem\cite{9} which reads ${
\chi
}_{1}(1)={\chi}_{3}(1)=0$.

At $\omega =1$ we find the following results in which we temporarily set
$g=h=0$.
\begin{equation} \begin{array}{c|cccc}{\frac{ \Lambda }{{  m}_{
q}}},n&{ \chi }_{1}(1)&{ \chi }_{2}(1)&{ \chi
}_{3}(1)&{ \xi }_{3}(1)\\  \hline
2,{\frac{3}{2}}&0&-0.153&0&
-0.002\\4,{\frac{3}{2}}&0&-0.132&0&-0.008\\2,
1&0&-0.148&0&-0.005\end{array} \label{table9} \end{equation}
\begin{equation} \begin{array}{c|cccc}{\frac{ \Lambda }{{  m}_{
q}}},n&{ \chi }_{1}^{ '}(1)&{ \chi }_{2}^{ '}(1)&{ \chi }_{3}^{ '}(1)&{
\xi }_{3}^{ '}(1)\\  \hline
2,{\frac{3}{2}}&-0.368&0.212&-
0.058&-0.097\\4,{\frac{3}{2}}&-0.307&0.161&-0.057&-
0.116\\2,1&-0.419&0.218&-0.073&-0.101\end{array}
\label{table10} \end{equation}

When $g$ and $h$ are nonzero the results in (\ref{table9}) do
not change.  The $g$ and $h$ dependence of various preceding quantities
has canceled out.  Only two of the first derivatives are changed by the
following amounts.
\begin{equation}
\delta {\chi }_{1}^{'}(1)=x(h-2g)
\label{a3} \end{equation}
\begin{equation}
\delta {\chi }_{3}^{'}(1)=-{\frac{1}{2}}xh
\label{a4} \end{equation}
$x=0.193,0.003,0.155$ for $(\Lambda /{m}_{q},n)$ = $(2,{\frac{3}{2}})$,
$(4,{\frac{3}{2}})$, and $(2,1)$ respectively.  In fact ${\chi }_{2}(\omega )$
and
${\xi }_{3}(\omega )$ are independent of $g$ and $h$ for any $\omega $,
as demonstrated in Appendix B.

We may translate these results into values for the ${\gamma }_{i}(1)$'s.  At
this stage
we are finally forced to make a choice for ${m}_{q}$.  We choose ${m}_{q}=250$
MeV as a representative value of an effective constituent quark mass in our
loops.
Using ${m}_{b}=4.8$ GeV, ${m}_{c}=1.44$ GeV and $g=h=0$ we obtain the
following results for ${ \gamma }_{i}(1)$ and ${ \gamma }_{i}^{ '}(1)$
expressed
as percentages of unity.
\begin{equation} \begin{array}{c|rrrrrr} \Lambda    , n& { \gamma
}_{+}(1) & { \gamma }_{- }(1)& { \gamma }_{V}(1) & { \gamma
}_{{A}_{1}}(1) & { \gamma }_{{A}_{2}}(1) & { \gamma }_{{A}_{3}}(1) \\
\hline
0.5\,{\rm GeV} , {\frac{3}{2}} &0& -8.5 &15.8&0&-19.5&11.1\\1.0\,{\rm
GeV} ,
{\frac{3}{2}}&0&-14.8&27.1&0&-31.6&17.5\\0.5\,{\rm GeV} , 1&0&-
9.7&17.9&0&-21.8&12.3\end{array} \label{table11} \end{equation}
\begin{equation} \begin{array}{c|rrrrrr} \Lambda    , n&  { \gamma
}_{+}^{ '}(1)  & { \gamma }_{-}^{ '}(1) & { \gamma }_{V}^{ '}(1)  & { \gamma
}_{{A}_{1}}^{ '}(1) & { \gamma }_{{A}_{2}}^{ '}(1) & { \gamma
}_{{A}_{3}}^{
'}(1) \\  \hline
0.5\,{\rm GeV} , {\frac{3}{2}} &-12.9&-1.7&-8.4&-1.2&8.4&-2.2\\1.0
\,{\rm GeV}
, {\frac{3}{2}}&-20.8&-3.6&-11.3&0.7&14.0&0.5\\0.5\,{\rm GeV} , 1&-
20.1&-
2.1&-11.2&-3.2&9.3&-3.9\end{array} \label{table12} \end{equation}
Again the only effect of nonzero $g$ and $h$ are in the following first
derivatives.
\begin{equation}
\delta {\gamma }_{V}^{'}(1)=\delta {\gamma }_{{A}_{1}}^{'}(1)=\delta
{\gamma }_{{A}_{3}}^{'}(1)=-2\overline{\Lambda }x\left({{\frac{g-
h}{{m}_{c}}}+{\frac{g+h}{{m}_{b}}}}\right)
\label{a1} \end{equation}
\begin{equation}
\delta {\gamma }_{+}^{'}(1)=-2\overline{\Lambda
}x(h+g)\left({{\frac{1}{{m}_{c}}}+{\frac{1}{{m}_{b}}}}\right)
\label{a2} \end{equation}
$x$ takes the same values as above.

An important point is that the zero recoil values of the four universal
functions in
(\ref{table9}) and the corresponding first order corrections to the form
factors in
(\ref{table11}) are independent of hyperfine mass splitting, as modeled by $g$
and
$h$.  Another is the weak dependence of the results in (\ref{table9}) and
(\ref{table10}) on the parameters $n$ and $ \Lambda $.  The results in
(\ref{table11}) and (\ref{table12}) show more variation and reflect the
fact that they are proportional to $\overline{\Lambda }$.  Corresponding to
the three rows of these tables we have $\overline{\Lambda }\approx$ 350,
600, 400 MeV respectively.

In our discussion thus far we have been more concerned with the sensitivity of
the
results to the parameters rather than trying to find an optimal set of
parameters.  The
latter could be accomplished in the following way.  We may calculate a
${\Lambda
}_{+}$ for each of ${B}^{{}^*}$ and ${D}^{{}^*}$ and a ${\Lambda }_{-}$ for
each of $B$ and $D$ by fitting the zeros of the full mass functions to the
physical meson masses.  By fitting these four ${\Lambda }_{\pm }$'s to the
first order form in (\ref{eq6}) we find, for $n=1$, $\Lambda=667$ MeV,
$g=0.057$, $h=0.37$, and for $n=3/2$, $\Lambda=818$ MeV, $g=0.047$,
$h=0.32$.  In both cases $\overline{ \Lambda } =504$ MeV which coincides
with a sum rule estimate\cite{3,7}.

Our results for the corrections are somewhat different from a sum rule
calculation\cite{3} which gives ${ \xi }_{3}(1)=0.33$ and ${ \chi }_{2}(1)=0$
and
from an improved sum rule calculation\cite{11} which gives ${ \chi }_{2}(1)=-
0.038$.  Note that in QCD ${ \chi }_{2}(1)$ measures the effect of a
chromo-magnetic moment operator insertion.\cite{9}  Our values for ${ \gamma
}_{i}(1)$ and ${ \gamma }_{i}^{ '}(1)$ may be compared directly with the
corresponding results in \cite{3}.

In this paper we have presented the zeroth and first order model results
for various quantities of interest in the meson heavy quark effective
theory.  The same model allows the calculation of physical quantities
to all orders in the $1/{m}_{Q}$ expansion.  This comparison, to pursued
elsewhere,\cite{15} should shed further light on the usefulness of the
heavy quark quark expansion in $B$ and $D$ meson physics.
\vspace{2ex}

\noindent {\bf Appendix  A}
\vspace{2ex}

We treat in this Appendix the zeroth order pieces of (\ref{eq7}-\ref{eq9}).
\vspace{1ex}

\noindent {\bf 1.  Masses and normalizations}
\vspace{1ex}

We begin with the pseudoscalar and vector mass functions
$\Gamma_{P,V}(p^{2})$ defined by setting the self-energy graphs equal to
$i\Gamma_{P}$ and $-ig_{\mu\nu}\Gamma_{V}+\cdots$ respectively,
where the ellipsis denotes $p_{\mu}p_{\nu}$ terms. The
light quark momentum $k$ may be chosen to be the same as the loop momentum,
and the heavy quark momentum is then $vM_{P,V}+k$ where $v^{2}=1$.
In the heavy quark limit we may ignore $k$ compared with $M_{P,V}$ and use
(\ref{eq8}) to show that the heavy
quark propagator becomes (after scaling $k\rightarrow \Lambda k$)
\begin{equation}
\frac{-i}{\Lambda}\frac{v\!\!\!/+1}{-2k\cdot v -c_{P,V}}.
\label{eqA1} \end{equation}
We have temporarily allowed the constants $c_{P,V}$ describing the
lowest order difference between quark and meson mass to be different for
pseudoscalar and vector.  Our first task is to show that they are in
fact equal to a common constant $c$.

The relevant traces are
\begin{equation}
{\rm Tr}\,
\left\{ \begin{array}{c} \gamma_{5} \\ -i\gamma_{\mu} \end{array} \right\}
(v\!\!\!/+1)
\left\{ \begin{array}{c} \gamma_{5} \\ -i\gamma_{\nu} \end{array} \right\}
(k\!\!\!/+\alpha)
\label{eqA2} \end{equation}
where $\alpha=m_{q}/\Lambda$. The translation is $k\rightarrow k-xv$
where $x$ is a Feynman parameter; the integral linear in $k$ then
vanishes.  We may anticommute the resulting $v$ leftwards past the
gamma matrices since terms proportional to $v_{\nu}$ do not contribute
in the vector case.  The results are
\begin{equation}
(x+\alpha){\rm Tr}\,
\left\{ \begin{array}{c} \gamma_{5} \\ -i\gamma_{\mu} \end{array} \right\}
(v\!\!\!/+1)
\left\{ \begin{array}{c} \gamma_{5} \\ -i\gamma_{\nu} \end{array} \right\}
=\overline{\xi}
\left\{ \begin{array}{c} 4 \\ -4g_{\mu\nu} \end{array} \right\},
\label{eqA3} \end{equation}
where $\overline{\xi}\equiv -k\cdot v + \alpha$ is equivalent with
$x+\alpha$.  We thus find
\begin{equation}
\Gamma_{P,V}(M_{P,V}^{2})=\int{\frac{d^{4}k}{(2\pi)^{4}}}
\frac{4in_{c}Z_{P,V}^{4n}\Lambda^{2-4n}}{[-{k}^{2}+1]^{2n}}
\frac{\overline{\xi}}{[-k^{2}+\alpha^{2}][-2k\cdot v-c_{P,V}]}  ,
\label{eqA4} \end{equation}

We are regarding these mass functions as the full two-point functions of the
mesons;
$i.e.$ there are no tree level contributions.  Thus the meson masses are
defined
by the locations of the zeros of the real parts of these mass functions:
\begin{equation} {\rm Re} \; \Gamma_{P,V}(M_{P,V}^{2})=0
.\label{eqA5} \end{equation}
Notice that imaginary parts are present in any quantity which is evaluated on
meson
mass shell.  This is because the difference  $\overline{ \Lambda }$ between the
meson
mass and the heavy quark mass is greater than the light quark mass ${m}_{q}$,
and
thus the meson is above the threshold for two free quarks.  But note that our
$\overline{ \Lambda }$ is  consistent with that of other approaches.\cite{3,7}
We
will not consider these imaginary parts further.  We also find and then drop an
overall, physically irrelevant minus sign.\cite{15}

We regard $c_{P}$ and $c_{V}$ as variables whose value is fixed by
(\ref{eqA4}) and (\ref{eqA5}).  It is then obvious that they are equal to
a common value $c$ because the functions $\Gamma_{P,V}(M_{P,V}^{2})$ differ
at most by multiplicative factors.  This demonstrates that ${M}_{P}={M}_{V}$ at
zeroth order (see (\ref{eq8})).

The mass functions are also required to satisfy the
normalization condition
\begin{equation}
\Gamma_{P,V}^{'}(M_{P,V}^{2})=1  .
\label{eqA6} \end{equation}
This
immediately implies that the normalization factors $Z_{P,V}$ are equal
to one another in the heavy quark limit.  The value of the zeroth order
constant $A$ defined in (\ref{eq7}) is thus fixed by
\begin{equation} 1=
\frac{\overline{\xi}}{D_{k}^{2n}D_{\alpha}D^{2}} ,\label{eqA7}
\end{equation}
where
$D_{k}=-k^{2}+1$, $D_{\alpha}=-k^{2}+\alpha^{2}$, $D=-2k\cdot v -c$.
Here and in the following we will adopt the convention that an overall
factor $4in_{c}A^{4n}$ and $\int d^{4}k/(2\pi)^{4}$ are understood
whenever the denominator is written explicitly as a product of $D$'s.

By the Ward identity the above determination of $A$ is equivalent to the
normalization of the Isgur-Wise function.  This will be verified explicitly
below.
\vspace{1ex}

\noindent {\bf 2.  Decay Constants}
\vspace{1ex}

We again choose the light quark momentum to coincide with the loop
momentum.  The traces relevant to the decay constants as defined by
(\ref{eq1and2}) at lowest order are
\begin{eqnarray}
{\rm Tr}\,
\left\{ \begin{array}{c} \gamma_{\mu}\gamma_{5} \\ \gamma_{\mu} \end{array}
\right\}
(v\!\!\!/+1)
\left\{ \begin{array}{c} \gamma_{5} \\ -i\gamma_{\nu} \end{array} \right\}
(k\!\!\!/+\alpha)
&=& \overline{\xi}\, {\rm Tr} \,
\left\{ \begin{array}{c} \gamma_{\mu}\gamma_{5} \\ \gamma_{\mu} \end{array}
\right\}
(v\!\!\!/+1)
\left\{ \begin{array}{c} \gamma_{5} \\ -i\gamma_{\nu} \end{array} \right\}
\nonumber \\
&=& \overline{\xi}
\left\{ \begin{array}{c} -4v_{\mu} \\ -4ig_{\mu\nu} \end{array}
\right\}
\label{eqA9} \end{eqnarray}
Use of the lowest order pieces of the expansions (\ref{eq6}-\ref{eq8})
immediately shows that the decay constants are equal at lowest order and
are given by
\begin{equation}
f_{P,V}=\frac{\Lambda^{3/2}}{m_{Q}^{1/2}}\times\frac{-A^{-2n}\overline{\xi}}
{D_{k}^{n}D_{\alpha}D}
\label{eqA10} \end{equation}

\noindent {\bf 3.  Form factors}
\vspace{1ex}

We turn to the zeroth order results for the meson
form factors defined by (\ref{eq3}-\ref{eq5}).
The relevant traces are
\begin{equation}
{\rm Tr} \,
\left\{ \begin{array}{c} \gamma_{5} \\ -i\gamma_{\nu} \end{array}
\right\}
(v\!\!\!/_{2}+1)
\Gamma_{\mu}
(v\!\!\!/_{1}+1) \gamma_{5}
(k\!\!\!/+\alpha)
\label{eqA12} \end{equation}
The translation is $k \rightarrow k-xv_{1}-yv_{2}$ with $v_{i}^{2}=1$,
where $x$ and $y$ are Feynman parameters.  The term linear in $k$
vanishes, and $v_{1}$ and $v_{2}$ may be anticommuted to the left and
right respectively to yield the standard traces
\begin{equation}
(x+y+\alpha) {\rm Tr} \,
\left\{ \begin{array}{c} \gamma_{5} \\ -i\gamma_{\nu} \end{array}
\right\}
(v\!\!\!/_{2}+1)
\Gamma_{\mu}
(v\!\!\!/_{1}+1) \gamma_{5}
\label{eqA13} \end{equation}
Use of the lowest order pieces of the expansions (\ref{eq6}-\ref{eq8})
and evaluation of the standard traces immediately yields the relations
\begin{equation}
h_{+}(\omega) = h_{V}(\omega) = h_{A_{1}}(\omega) =
h_{A_{3}}(\omega) \equiv \xi(\omega) \;\; {\rm and} \;\;
h_{-}(\omega) = h_{A_{2}}
(\omega) =  0 \label{eqA17} \end{equation}

The quantity $x+y+\alpha$ is equivalent to
$\tilde{\xi}\equiv -(1+\omega)^{-1} k\cdot (v_{1}+v_{2})+\alpha$, and
we find for the Isgur-Wise function
\begin{equation} \xi=
\frac{\tilde{\xi}}{D_{k}^{2n}D_{\alpha}D_{1}D_{2}}
\label{eqA18} \end{equation}
where $D_{i}=-2k \cdot v_{i}-c$.
The normalization $\xi(1)=1$ is verified by setting $v_{1}=v_{2}=
v$ and comparing with (\ref{eqA7}).
\vspace{2ex}

\newpage
\noindent {\bf Appendix B}
\vspace{1ex}

The first order correction terms to the weak decay form factors (defined in
(\ref{eq10}) and (\ref{c1})) may be expressed in terms of the universal
functions as
follows.\cite{4,9}
\begin{eqnarray}
{ \gamma }_{+} \xi & = & \left( \frac{\overline{\Lambda}}{{ m}_{ b}}
 + \frac{\overline{\Lambda }}{{ m}_{ c}} \right)
({\chi }_{1} + 2[1 - \omega  ]{\chi }_{2} + 6{\chi }_{3}) \label{eqB1}  \\
 { \gamma }_{-} \xi  & = & \left(  \frac{\overline{\Lambda }}{{ m}_{ c}}
- \frac{\overline{\Lambda }}{{ m}_{ b}} \right)
\left( {\xi }_{3} - \frac{1}{2} \xi \right) \label{eqB2} \\
{ \gamma }_{V} \xi  & = & \frac{\overline{\Lambda }} {2{ m}_{ c}}
(\xi  +2{\chi }_{1}-4{\chi }_{3}) + \frac{\overline{\Lambda }}{2{ m}_{ b}}
( \xi  -2{\xi }_{3}+2{\chi }_{1}+4[1-\omega  ]{\chi }_{2}+12{\chi }_{3})
\label{eqB3} \\
{ \gamma }_{{A}_{ 1}} \xi &  = & \frac{\overline{\Lambda }}{2{ m}_{ c}}
\left(  \frac{\omega  -1}{\omega  +1} \xi  +2{\chi
}_{1}-4{\chi }_{3} \right)  \nonumber \\
&& + \frac{\overline{\Lambda }} {2{ m}_{ b}}
\left( \frac{\omega  -1}{\omega  +1} [\xi  -2{\xi }_{3}]+
2{\chi }_{1}+4[1-\omega  ]{\chi }_{2}+12{\chi }_{3} \right)
\label{eqB4} \\
{ \gamma }_{{A}_{ 2}} \xi & = & \frac{\overline{\Lambda }}{{ m}_{ c}}
\left( - \frac{1}{\omega  +1} [\xi  +{\xi }_{3}]+
2{\chi }_{2} \right) \label{eqB5} \\
 { \gamma }_{{A}_{ 3}} \xi & = & \frac{\overline{\Lambda }}{2{ m}_{ c}}
\left( \frac{1}{\omega  +1} [\{\omega  -1\}\xi  -
2{\xi }_{3}]+2{\chi }_{1}-4{\chi }_{2}-4{\chi }_{3} \right)
\nonumber  \\
&& + \frac{\overline{\Lambda }}{2{ m}_{ b}}
(\xi  -2{\xi }_{3}+2{\chi }_{1}+4[1-\omega  ]{\chi
}_{2}+12{\chi }_{3}) \label{eqB6}
\end{eqnarray}
In the notation of (\ref{eq10}) and Appendix A we obtain
\begin{eqnarray}
\delta_{b}h_{i}&=&
  \alpha_{i}\left\{ (2B_{P}-\frac{c}{4})\xi
	+\frac{2(g+h)\tilde{\xi}}{D_{k}^{3}D_{\alpha}D_{1}D_{2}}
	+\frac{(k^{2}-\frac{c^{2}}{2}+cd_{P})\tilde{\xi}}
	{D_{k}^{2}D_{\alpha}D_{1}^{2}D_{2}} \right. \nonumber \\
& &  \mbox{} + \left. (\frac{\alpha}{2}+\frac{c}{4})
  \frac{(1+\omega)^{-1}k\cdot(v_{1}+v_{2})}
	{D_{k}^{2}D_{\alpha}D_{1}D_{2}}
	\right\}
	+\frac{X_{i}}{D_{k}^{2}D_{\alpha}D_{1}D_{2}}
\label{bird} \end{eqnarray}
\begin{eqnarray}
\delta_{c}h_{i}&=&
  \alpha_{i}\left\{ (2B_{P,V}-\frac{c}{4})\xi
	+\frac{2(g\pm h)\tilde{\xi}}{D_{k}^{3}D_{\alpha}D_{1}D_{2}}
	+\frac{(k^{2}-\frac{c^{2}}{2}+cd_{P,V})\tilde{\xi}}
	{D_{k}^{2}D_{\alpha}D_{1}D_{2}^{2}} \right. \nonumber \\
& &  \mbox{} + \left. (\frac{\alpha}{2}+\frac{c}{4})
  \frac{(1+\omega)^{-1}k\cdot(v_{1}+v_{2})}
	{D_{k}^{2}D_{\alpha}D_{1}D_{2}}
	\right\}
	+\frac{Y_{i}}{D_{k}^{2}D_{\alpha}D_{1}D_{2}}
\label{dog} \end{eqnarray}
where $P$ or $V$ quantities are used in the expression for $\delta_{c}h_{i}$
according to whether the final state is pseudoscalar or vector,
and $X_{i}$ and $Y_{i}$ are given by
\begin{equation}
\begin{array}{c|c|c|c}
i & \alpha_{i} & X_{i} & Y_{i} \\  \hline
+ & 1 & -\frac{1}{2}k^{2} -\frac{\alpha c}{4} & X_{+} \\
- & 0 & \frac{1}{2}k^{2}+\frac{\alpha c}{4} & -X_{-} \\
V & 1 & 0 & 0 \\
A_{1} & 1 & -\frac{1}{1+\omega}k^{2} -\frac{c\alpha}
{2(1+\omega)} & \frac{2}{1+\omega}S + X_{A_{1}} \\
A_{2} & 0 & 0 & -2T+\alpha(1+\omega)^{-1} k\cdot(v_{1}+v_{2}) \\
A_{3} & 1 & 0
    & -2U
\label{cat} \end{array} \end{equation}
The quantities $S$, $T$ and $U$ are defined by
\begin{equation}
\frac{k_{\mu}k_{\nu}}{D_{k}^{2}D_{\alpha}D_{1}D_{2}}=
\frac{Sg_{\mu\nu}
+T(v_{1\mu}v_{1\nu}+v_{2\mu}v_{2\nu})+U(v_{1\mu}v_{2\nu}+
v_{2\mu}v_{1\nu})}{D_{k}^{2}D_{\alpha}D_{1}D_{2}}
\end{equation}

On the other hand, proper normalization of the vector meson ``charge" graph at
$q^{2}=0$ gives in the notation of Appendix A
\begin{equation}
0= 2B_{V}-\frac{c}{4}
	+\frac{2(g-h)\overline{\xi}}{D_{k}^{3}D_{\alpha}D^{2}}
	+\frac{(k^{2}-\frac{c^{2}}{2}+cd_{V})\overline{\xi}}
	{D_{k}^{2}D_{\alpha}D^{3}}
	+(\frac{\alpha}{2}+\frac{c}{4})\frac{k\cdot v}
	{D_{k}^{2}D_{\alpha}D^{2}}
	+\frac{\overline{Y}_{V}}{D_{k}^{2}D_{\alpha}D^{2}}
\label{bob} \end{equation}
where
\begin{equation}
\overline{Y}_{V}=S-\frac{1}{2}k^{2}-\frac{c\alpha}{4}
\label{mark} \end{equation}
Setting $v_{1}=v=v_{2}$ in (\ref{dog}) and (\ref{cat}) and comparing
with (\ref{bob}) and (\ref{mark}) shows that $\delta_{c}h_{A_{1}}=0$
at $\omega=1$.

An analogous argument using the normalization of the pseudoscalar
meson charge graph shows that $\delta_{b}h_{A_{1}}=0$.
Luke's theorem, ${ \chi }_{1}(1)={ \chi}_{3}(1)=0$, is therefore satisfied for
arbitrary $m_{c}$ and $m_{b}$ with no extra constraints between model
parameters.

Returning to arbitrary $\omega $ we find that all the required relations which
are supposed to hold between the ${\delta }_{b}{h}_{i}$ follow
straightforwardly from the model.  The following two relations which need to
be checked for the $\delta_{c}{h}_{i}$ are less trivial.
\begin{equation}
\delta_{c}h_{V}-\delta_{c}h_{A_{1}}=\frac{c\xi}{2(1+\omega)}
\label{relation} \end{equation}
\begin{equation}
\delta_{c}(h_{A_{1}}-h_{A_{2}}-h_{A_{3}}-2(1+\omega)^{-1}h_{-})
=\frac{c\xi}{(1+\omega)}
\end{equation}
These are satisfied only if the following nontrivial identities are satisfied.
\begin{equation}
\frac{{\frac{2}{c}}S-\frac{1}{c}k^{2}-\frac{k\cdot
(v_{1}+v_{2})}{2(1+\omega)}}
{D_{k}^{2}D_{\alpha}D_{1}D_{2}}=0
\end{equation}
\begin{equation}
\frac{S+(1+\omega)(T+U)-\frac{\alpha}{2}k\cdot(v_{1}+v_{2})}
{D_{k}^{2}D_{\alpha}D_{1}D_{2}}=\frac{c}{2}\xi
\end{equation}
We have confirmed these relations numerically.

The functions $\chi_{2}$ and ${\xi }_{3}$ are independent of $g$ and $h$.  We
find
\begin{equation}
\xi_{3}=\frac{-\frac{1}{c}k^{2}-\frac{k\cdot (v_{1}+v_{2})}{2(1+\omega)}}
{D_{k}^{2}D_{\alpha}D_{1}D_{2}}
\end{equation}
\begin{equation}
\chi_{2}=\frac{\xi+\xi_{3}}{2(1+\omega)}+
\frac{-\frac{2T}{c}+\frac{\alpha}{c}
\frac{k\cdot(v_{1}+v_{2})}{1+\omega}}
{D_{k}^{2}D_{\alpha}D_{1}D_{2}}.
\end{equation}
\vspace{2ex}

\noindent {\bf Acknowledgment}
\vspace{1ex}

This research was supported in part by the Natural Sciences and
Engineering Research Council of Canada.


\end{document}